# UNCONDITIONALLY SECURE CREDIT/DEBIT CARD CHIP SCHEME AND PHYSICAL UNCLONABLE FUNCTION

LASZLO B. KISH [(1)], KAMRAN ENTESARI [(1)], CLAES-GÖRAN GRANQVIST [(2)], CHIMAN KWAN [(3)]

[(1)] Department of Electrical and Computer Engineering, Texas A&M University, College Station, TX 77843-3128, USA; email: Laszlo.Kish@ece.tamu.edu

[(2)] Department of Engineering Sciences, The Ångström Laboratory, Uppsala University, P.O. Box 534, SE-75121 Uppsala, Sweden

[(3)] Signal Processing Inc., 9605 Medical Center Dr., Rockville, MD 20850, USA

**Abstract.** The statistical-physics-based Kirchhoff-law–Johnson-noise (KLJN) key exchange offers a new and simple unclonable system for credit/debit card chip authentication and payment. The key exchange, the authentication and the communication are unconditionally secure so that neither mathematics- nor statistics-based attacks are able to crack the scheme. The ohmic connection and the short wiring lengths between the chips in the card and the terminal constitute an ideal setting for the KLJN protocol, and even its simplest versions offer unprecedented security and privacy for credit/debit card chips and applications of physical unclonable functions.

## 1. Introduction

Modern chip-based credit, debit and bank cards use physical unclonable function (PUF) hardware keys for payment, and their protocol and cryptography employ the Europay, MasterCard and Visa (EMV) scheme [1]. Various levels of strengths for PUF hardware keys are described in the literature [2]. The PUF in these cards is weak because it relies on *conditionally* secure communication, data authentication and a private key stored in a secure memory that must be able to self-destruct when tampered with. Thus the cards' software-based secure communication protocols (such as RSA [1]) offer nothing but conditional security, and it may not be surprising that criminals have been able to crack the EMV system and steal large amounts of money [1]. In fact, the hacking referred to above [1] was even simpler and utilized simple vulnerabilities caused by an incomplete security and authentication protocol (including a man-in-the-middle-attack).

For systems with *complete* conditional security and authentication protocols, though it appears unlikely that a hacker has sufficient computational power or is able to invent a sufficiently efficient factoring algorithm to crack the key, a skilled hacker can efficiently employ a *known-plaintext attack*, which constitutes the major vulnerability of all conventional mathematics-based secure key exchange protocols [3].

For high-security credit cards and PUF applications, it is essential to use *unconditionally secure communication* [3] which is immune also against known-plaintext attacks; otherwise the cards' entire PUF protocol is highly vulnerable and hackers will no doubt in the foreseeable future develop devices to copy them. Unconditionally secure key exchange with *one-time-pad* (OTP) encryption makes the communication unconditionally secure and immune against any type of attack, including mathematics-based and statistics-based alternatives.

Up till now, only two schemes offer unconditional security: (*i*) quantum encryption including its enhanced versions [3] and (*ii*) the electronic-noise-based Kirchhoff-law–Johnson-noise (KLJN) scheme [4] which is the main topic of the present article and is elaborated below. A disadvantage of quantum encryption is that it is bulky and hence it cannot be integrated on a chip. KLJN, on the other hand, offers the possibility of chip integration. The latter scheme requires electrical wire connection between the communicating parties, which makes it well suited for chip-based credit and debit cards which employ wire connection to communicate with the terminal during payment, and consequently a KLJN chip is feasible for applications.

In Section 2, we briefly describe the KLJN scheme, which is the core of the unconditionally secure credit card protocol of unclonable credit/debit cards and PUF applications discussed in Section 3.

## 2. The KLJN key exchange scheme

The KLJN secure key exchange scheme [4–13] was proposed in 2005 [5,6] and, including its advanced versions [8–10], there are currently about ten different protocols. It is a statistical/physical competitor to quantum key exchange, and the security is based on Kirchhoff's Loop Law and the Fluctuation–Dissipation Theorem. The security of the ideal KLJN scheme is as strong as the impossibility to build a perpetual motion machine of the second kind.

Figure 1 depicts a binary version of the KLJN scheme and shows that, during a single-bit exchange, the communicating parties (Alice and Bob) connect their randomly chosen resistor (including its Johnson noise generator) to a wire channel. These resistors are randomly selected from the publicly known set $\{R_L, R_H\}$, $R_L \neq R_H$, representing the Low and High bit values. The Gaussian voltage noise generators—mimicking the Fluctuation–Dissipation Theorem and delivering band-limited white noise with publicly agreed bandwidth—produce enhanced thermal (Johnson) noise at a publicly agreed effective temperature $T_{\text{eff}}$; typically $T_{\text{eff}} \gg 10^{10}\,\text{K}$ so that the temperature of the wire can be neglected. The noises are statistically independent of each other and from the noise of the former bit period.

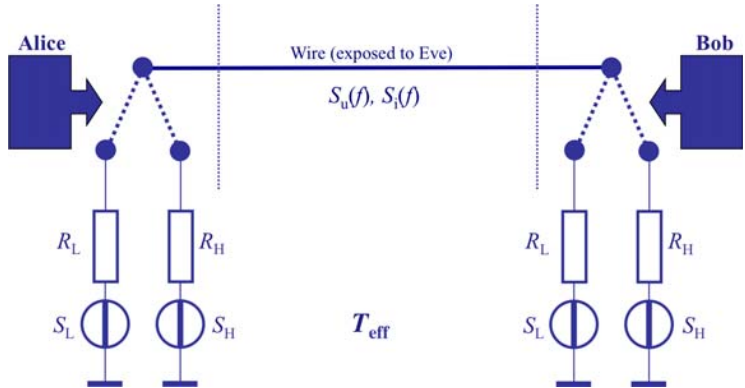

Fig. 1. Core of the KLJN scheme without defense circuitry [4,5] against active (invasive) attacks and attacks utilizing non-idealities. The $R_L$ and $R_H$ resistors, identical pairs at Alice and Bob, represent the Low (*L*) and High (*H*) bit values. The corresponding noise spectra $S_L$ and $S_H$ also form identical pairs at the two ends, but they belong to independent Gaussian stochastic processes. Both parties are at the same temperature, and thus the net power flow is zero. The *LH* and *HL* bit situations of Alice and Bob produce identical voltage and current noise spectra, $S_u$ and $S_i$, in the wire, implying that they represent a secure bit exchange. The *LL* and *HH* bit arrangements, which occur in 50% of the cases, have singular noise levels in the wire, and hence they do not offer security because Eve can distinguish them. Consequently 50% of the bits must be discarded. This system works also with arbitrary, non-binary resistor values as an analog circuitry to exchange continuum information about the distribution of random resistors.

In the case of secure bit exchange—*i.e.*, the *LH* or *HL* bit situations for Alice and Bob—an eavesdropper (Eve) cannot distinguish between these two situations by measuring the noise spectra, $S_u(f)$ and $S_i(f)$, of voltage and/or current in the cable, respectively, because the *LH* and *HL* noise levels are identical (degenerated). Thus when Alice and Bob detect the noise spectra (or noise levels) characteristic of the *LH* and *HL* situation, they know that the other party has the opposite bit and that this bit is secure. Then one of them will invert the bit (it is publicly pre-agreed who will do this) to get the same key bit as the other party. The KLJN scheme offers unconditional (information theoretic) security under both ideal and slightly non-ideal (*i.e.*, practical) conditions [4].

The security against *active* (invasive) *attacks* is provided by the robustness of classical-physical quantities, which guarantees that they can be continuously monitored and exchanged between Alice and Bob via authenticated communication [11,12]. Therefore the system, and the consistency of the measured and exchanged voltage and current data with known cable parameters and model, can be checked continuously and deterministically without destroying these data, which is totally different from the case of a quantum key distribution. The same defense method provides natural immunity against the *man-in-the-middle attack* [11], which was part of the successful credit card attack described before [1].

One should keep in mind that the KLJN secure information exchanger is basically an analog circuit and can work with arbitrary resistances because, even if the resistance values are not pre-agreed, Alice can calculate Bob's resistance from the measured data [5] by using Johnson's formula, and *vice versa*. For example, by using the measured current spectrum in the wire one obtains

$$R_\text{B} = \frac{4kT_\text{eff}}{S_i} - R_\text{A} \ , \tag{1}$$

where *k* is Boltzmann's constant. It is important to note that Eve is also able to determine an arbitrary, non-pre-agreed (non-publicly known) resistor pair connected to the line by using measured voltage and current spectra [5]. The two solutions of the relevant second order equation provide two resistance values of the pair according to

$$R_{1,2} = \frac{4kTS_u \pm \sqrt{(4kTS_u)^2 - 4S_u^3 S_i}}{2S_u S_i} \ . \tag{2}$$

However, Eve cannot determine which resistor is with Alice and which is with Bob, and hence the information exchange about the distribution of arbitrary, non-binary resistor values is secure in the original KLJN system.

Under ideal conditions, the system delineated above offers unconditional (information theoretic) security against passive (listening) attacks only. To achieve unconditional security against active (invasive) attacks, and/or attacks that utilize non-ideal features of the building elements, the following extra elements are essential: (*a*) an external Gaussian noise generator to emulate thermal noise [5] and to have stable noise temperature with high accuracy, (*b*) filters for limiting the frequency range to the measurement bandwidth and to the no-wave limit [5], and (*c*) measurement of the instantaneous voltage and current values at the two ends (by Alice and Bob) and comparison of them via an authenticated public channel [11] or, in more advanced set-ups, with a cable simulation model fed by real-time data [12]. A difference between the two measurements in (*c*) serves as an indication of invasive eavesdropping, and the corresponding bits must be discarded. Section 3 below discusses the number of secure bits consumed by the authentication.

Several active attacks have been proposed against the KLJN system [5,11,12], but the defense method mentioned above is able to protect the system and maintain its unconditional security under general conditions [12]. Specifically, the KLJN system has been proposed to deliver unconditional security for hardware in computers, games and instruments [13] and to be used as a PUF device [2].

## 3. The new credit/debit/PUF card protocol

The credit/debit/PUF card (referred to as the "Card" below) is used by the consumer at the terminal where the Card is plugged in, and a wire connection is then built between the chip in the Card and the interface in the terminal as indicated in Figure 2. The terminal interface communicates with a server (located at banks, etc.), which is not shown in this figure. We furthermore assume that the communication between the terminal and the server is secure and free from malevolent parties. The protocol entails several phases as elaborated in detail below.

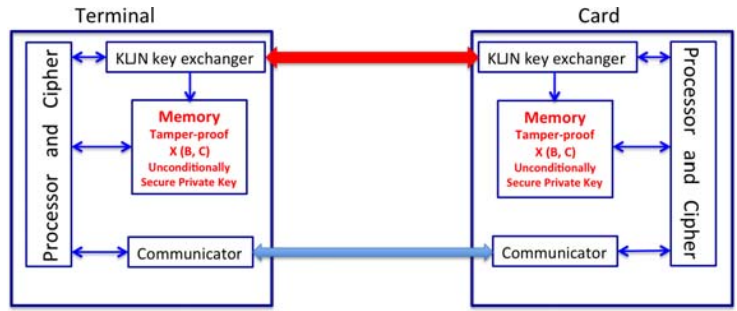

Fig. 2. Core hardware scheme of the terminal and the Card chip. Units regarding terminal–server communication, transaction input, etc., are not shown.

### 3.1. *Private key initialization during fabrication*

During initialization, a private key—which is a binary bit-string *C*—is generated by a true random-number generator. It will be used for the authentication of current and voltage data during the KLJN key exchange and also for the authentication of the Card.

There is another memory location for a separate private key, *B*, which is used for the encryption; that memory section is left empty during fabrication. The private key *C* is stored on the server at a memory location that is associated with the publicly known card identification data (the public key), such as card number, cardholder's name, expiration date, etc. The private key is also stored in the tamper-proof secure memory on the Card.

Below, we use the term "session" to signify the *entire* card transaction process, including the "authentication" and the actual "transaction".

### 3.2. *Authentication of the Card*

During its use, when the Card is inserted into the terminal, the authentication of the Card takes place according to the steps below:

(*i*) the Card sends its publicly known card identification data (the public key) to the processor in the terminal, which forwards these data to the server;

(*ii*) the server locates the corresponding private key in its memory and forwards that information to the processor in the terminal;

(*iii*) the Card and terminal execute a KLJN key exchange;

(*iv*) authentication of the Card takes place during the verification of the current and voltage data, and the comparison will indicate a discrepancy and thus an active attack if an incorrect private key is used; the communication is then terminated, the session is broken, and the fact that an attack has taken place is recorded. The card is canceled after a certain small number *M* of broken sessions. It is essential that the private key *C* contains enough non-overlapping (independent) bits to cover the (*M* – 1) broken sessions and the additional procedure, which is either another broken session or a successful session.

### 3.3. *Transaction*

During the transaction, the Card and the terminal encrypt their communication by the newly-generated private key *B*. This communication includes also the identification of the user (via a PIN code), if needed. This step should also use an OTP cipher to provide unconditional security, particularly for the known-plain-text parts of the communication. After the transaction, the key *B* is deleted.

### 3.4. *Generating and sharing the new and unconditionally secure authentication key*

During each successful use of the Card, a new and independent unconditionally secure private key *C* is generated and is recorded for the next card session. As a consequence of the short distance (< 0.1 meter) between the key chips in the Card and the terminal, the conditions are practically ideal and the key exchange can take place at high speed and with minimal information leak towards passive (listening) attacks. Nevertheless, a simple three-stage XOR-based privacy amplification [14] (yielding an eightfold slowdown) may be applied to account for unforeseen situations.

The private key *C* is used for authentication of current and voltage data during the KLJN key exchange. Assuming simple hashing-based [15] data authentication, the minimum length $N_C$ of *C* is of the order of

$$N_C \approx M \log_2(N_d) ,\qquad(3)$$

where $N_d$ is the length of the string containing the current- and voltage-sampling data during a successful session and *M* was defined above.

### *3.5 Speed of the protocol*

Considering the very short wire separation between the Card chip and the chip in the terminal, one can conclude that the practical key exchange speed for the KLJN scheme is not limited by distance [5] but by electronics. Inexpensive chip design allows 100-kHz bandwidth in an effortless way, which translates to a KLJN key exchange speed of about thousand bits/second. This rate guarantees that the generation/sharing of the new KLJN key, which is the most time consuming process of the protocol, takes only two to three seconds even if long private keys (such as 1024-bits) are used.

### 4. Conclusion

We showed that the KLJN key exchange scheme offers an unconditionally secure, simple and unclonable system for credit/debit card chip authentication and payments. The key exchange, the authentication and the communication are unconditionally secure, and thus neither mathematics- nor statistics-based attacks are able to crack the scheme. The proposed simple protocol offers unprecedented security for credit/debit card chips and PUF applications by providing new, unconditionally secure private keys after each transaction. The projected costs of the proposed unconditionally secure chip are of the same order of magnitude as the cost of today's conditionally secure credit card chips.

## Acknowledgements

We are grateful to Gergely Vadai and Robert Mingesz for critical remarks and discussions, particularly for pointing out a redundant section in our former protocol that was presented in the first version of the manuscript.

## References

[1]  H. Ferradi, R. Géraud, D. Naccache, A. Tria, "When organized crime applies academic results: a forensic analysis of an in-card listening device", *J. Cryptogr. Eng*. **6** (2016) 49–59.
[2]  L.B. Kish, Ch. Kwan, "Physical unclonable function hardware keys utilizing Kirchhoff-law–Johnson-noise secure key exchange and noise-based logic", *Fluct. Noise Lett.* **12** (2013) 1350018.
[3]  H. Yuen, "Security of quantum key distribution", *IEEE Access* **4** (2016) 724–749.
[4]  L.B. Kish, C.G. Granqvist, "On the security of the Kirchhoff-law–Johnson-noise (KLJN) communicator", *Quant. Inform. Proc.* **13** (2014) 2213–2219.
[5]  L.B. Kish, "Totally secure classical communication utilizing Johnson(-like) noise and Kirchhoff's law", *Phys. Lett. A* **352** (2006) 178–182.
[6]  A. Cho, "Simple noise may stymie spies without quantum weirdness", *Science* **309** (2005) 2148. http://www.ece.tamu.edu/~noise/news_files/science_secure.pdf ; D.J. Palmer, "Noise encryption keeps spooks out of the loop", *New Scientist*, issue *2605* (2007) 32.
 http://www.newscientist.com/article/mg19426055.300-noise-keeps-spooks-out-of-the-loop.html .
[7]  R. Mingesz, Z. Gingl, L.B. Kish, "Johnson(-like)–noise–Kirchhoff-loop based secure classical communicator characteristics, for ranges of two to two thousand kilometers, via model-line", *Phys. Lett. A* **372** (2008) 978–984.
[8]  L.B. Kish, "Enhanced secure key exchange systems based on the Johnson-noise scheme", *Metrol. Meas. Syst.* **20** (2013) 191–204.
http://www.degruyter.com/view/j/mms.2013.20.issue-2/mms-2013-0017/mms-2013-0017.xml
[9]  G. Vadai, R. Mingesz, Z. Gingl, "Generalized Kirchhoff-law–Johnson-noise (KLJN) secure key exchange system using arbitrary resistors", *Sci. Rep.* **2015** (2015) 13653. doi:10.1038/srep13653.
[10] L.B. Kish, C.G. Granqvist, "Random-resistor–random-temperature Kirchhoff-law–Johnson-noise (RRRT–KLJN) key exchange", *Metrol. Meas. Syst.* **23** (2016) 3–11.
http://www.degruyter.com/view/j/mms.2016.23.issue-1/mms-2016-0007/mms-2016-0007.xml
[11] L.B. Kish, "Protection against the man-in-the-middle-attack for the Kirchhoff-loop–Johnson(-like)-noise cipher and expansion by voltage-based security", *Fluct. Noise Lett.* **6** (2006) L57–L63.
[12] H.P. Chen, M. Mohammad, L.B. Kish, "Current injection attack against the KLJN secure key exchange", *Metrol. Meas. Syst.***23** (2016) 173–181.
[13] L.B. Kish, O. Saidi, "Unconditionally secure computers, algorithms and hardware", *Fluct. Noise Lett.* **8** (2008) L95–L98.
[14] T. Horvath, L.B. Kish, J. Scheuer, "Effective privacy amplification for secure classical communications", *EPL* **94** (2011) 28002.
[15] W.B. Frakes, R. Baeza-Yates, eds., Information Retrieval: Data Structures and Algorithms, Prentice-Hall, New York (1992).